\documentclass[a4paper,11pt]{article}
\pdfoutput=1
\usepackage{jheppub} 

\usepackage[normalem]{ulem}
\usepackage[T1]{fontenc} 
\usepackage{subcaption}

\newcommand{\mzprime}{$m_{Z^{\prime}}$\,}
\newcommand{\zprime}{$Z^{\prime}$\,}
\newcommand{\gprime}{$g_1^{\prime}$\,}

\title{\boldmath Heavy neutrino production via $Z^{\prime}$ at the lifetime frontier}

\author[a, b]{Frank F. Deppisch,}
\author[b]{Suchita Kulkarni,}
\author[a]{Wei Liu}

\affiliation[a]{University College London,\\Gower Street, London WC1E 6BT, UK}
\affiliation[b]{Institut f{\"u}r Hochenergiephysik, {\"O}sterreichische Akademie der Wissenschaften, 
	Nikolsdorfer Gasse 18, 1050 Wien, Austria}

\emailAdd{f.deppisch@ucl.ac.uk}
\emailAdd{suchita.kulkarni@oeaw.ac.at}
\emailAdd{wei.liu.16@ucl.ac.uk}

\abstract{We investigate the pair production of right-handed neutrinos from the decay of an additional neutral $Z^{\prime}$ boson  in the gauged $B-L$ model. Taking into account current constraints on the $Z^{\prime}$ mass and the associated gauge coupling $g_{1}^{\prime}$, we analyse the sensitivity of proposed experiments at the lifetime frontier, FASER 2, CODEX-b, MATHUSLA as well as a hypothetical version of the MAPP detector to a long lived heavy neutrino $N$ originating in the decays of the $Z^{\prime}$. We further complement this study with determining the reach of LHCb and a CMS-type detector for the high-luminosity LHC run. We demonstrate that in a  background free scenario with $g_1^\prime = 10^{-3}$ near the current limit, FASER 2 is sensitive to the active-sterile neutrino mixing down to $V_{\mu N} \approx  10^{-4}$, while a reach of $V_{\mu N} \approx 10^{-5}$ can be obtained for CODEX-b and LHCb, in a mass regime of $m_N \approx 5-20$~GeV and $m_{Z^{\prime}} \approx 20-70$~GeV. Finally, MATHUSLA can probe $V_{\mu N} \approx 10^{-7}$ and cover the mixing regime expected in a canonical seesaw scenario of light neutrino mass generation.}
\begin{document} 
\maketitle
\flushbottom
\section{Introduction}
The observation of light neutrino mass from neutrino oscillation experiment shows the existence of physics beyond the Standard Model (SM). In order to explain the light neutrino mass, a simple ultraviolet (UV) complete model $U(1)_{B-L}$~\cite{Mohapatra:1980qe} can be written down where in addition to the Standard Model (SM) fermion content, three right-handed Majorana neutrinos are added. It contains a SM singlet Higgs $\chi$ which spontaneously breaks the $B-L$ symmetry and gives a Majorana mass for the heavy right-handed (RH) neutrinos $N_{i}$. 

This model provides an additional gauge boson $Z^{\prime}$ corresponding to the $B-L$ gauge  symmetry which can later decay into pairs of right-handed neutrinos generating lepton number violating processes and other SM fermions. In the small active-sterile neutrino mixing limit, this leads to distinctive displaced vertex signatures at the LHC. Such a process is controlled by the $B-L$ coupling  $g_1^{\prime}$, the mass of the heavy gauge boson ($Z^{\prime}$) and that of the heavy neutrinos ($N_i$). Such a heavy gauge boson can be searched for in several different ways. LHC searches for heavy resonance in dilepton final states put a strict bound on $m_{Z^{\prime}}>$ 4.5 TeV~\cite{Aaboud:2017buh} for a $Z^{\prime}$ with couplings similar to the SM $Z$ boson. While the $B-L$ breaking scale $x = M_{Z^\prime}/2g_1^\prime$ is constrained to be larger than 3.45~TeV from LEP-II~\cite{Heeck:2014zfa, Cacciapaglia:2006pk, Anthony:2003ub, LEP:2003aa, Carena:2004xs}, these limits fail when $m_{Z^\prime}$ becomes too small as the effective Lagrangians can not be applied. Limits on the full $M_{Z^\prime} -  g_1^{\prime}$  parameter space are available based on the LHC results using the Constraints On New Theories Using Rivet ({\tt CONTUR}) method for $g_1^{\prime} < 10^{-3}$ approximately for a $Z^{\prime}$ above 1~GeV~\cite{Amrith:2018yfb, Butterworth:2016sqg}, or even smaller for $g_1^{\prime} < 10^{-4}$ for $m_{Z^{\prime}} < 10$ GeV, and $g_1^{\prime} < 10^{-8}$ for $m_{Z^{\prime}} < 1$~GeV from recasting dark photon searches using {\tt Darkcast}~\cite{Ilten:2018crw}. With respect to the heavy neutrinos $N_i$ in this model, their lifetime crucially depends on their mixing $V_{l N}$ with the active light neutrinos. In a canonical Type-I seesaw mechanism of neutrino mass generation, the mass scale $m_\nu$ of light neutrinos is related to that of the heavy neutrinos as $m_{\nu} \sim V_{lN}^2 m_N$. As we know that the light neutrinos have a sub-eV mass scale, with constraints coming from $0\nu\beta\beta$ and Tritium beta decay experiments \cite{Kraus:2004zw, Aseev:2011dq} as well as from cosmological observations such as Planck \cite{Ade:2015xua}, the corresponding neutrino mixing is expected to be very small, $V_{l N} \approx 10^{-5} - 10^{-6}$, for heavy neutrino mass of order $m_N \approx 10 - 100$~GeV. The corresponding decay lengths are of the order of centimetres to metres, thus models of neutrino mass generation around the electroweak scale clearly motivate searches at the lifetime frontier. Note however that the relation between light and heavy neutrino mass can be modified in extended scenarios. For example in the inverse seesaw mechanism, the neutrino mixing can be large, only limited by experimental limits on the heavy neutrinos, see e.g. \cite{Deppisch:2015qwa} and references therein.

Searches for displaced vertices at the LHC and in dedicated experiments have gained significant attention. Several CMS and ATLAS analyses \cite{CMS:2014hka, CMS:2015pca, CMS:2014mca} have studied displaced vertices from long-lived neutral particles in various mass ranges, with no significant number of signal events observed so far. Specifically, direct searches for heavy neutrinos, which includes searches for displaced vertex signatures, have been studied in Refs.~\cite{Khachatryan:2015gha, Sirunyan:2018mtv, Aad:2019kiz}, in the context of sterile heavy neutrinos, i.e. where the $N_i$ only couple via their mixing with light neutrinos and there is no extra gauge interaction. These searches typically yield constraints of the order $V_{lN} \lesssim 10^{-3}$ \cite{Aad:2019kiz} for $m_N \lesssim m_Z$.

Apart from the popular dilepton final states, \zprime in $B-L$ models can also decay to pair of heavy neutrinos if kinematically accessible. In the regime of suppressed active-sterile neutrino mixing the decays of the right-handed neutrinos can lead to a displaced vertex signal at colliders. This distinct signal can be regarded as background free for large decay lengths, thus even under the highly constrained parameter space, the channel $pp \to Z^\prime \to N \ N$ leading to displaced vertices is a promising channel to search for at the proposed far detectors such as MATHUSLA~\cite{Chou:2016lxi}, MAPP~\cite{Pinfold:2019nqj}, CODEX-b~\cite{Gligorov:2017nwh}, FASER~\cite{Ariga:2018uku} as well as LHCb~\cite{Aaij:2016xmb} and CMS~\cite{CMS:2014hka, CMS:2015pca}. 

A number of similar reinterpretation studies have already been made to search for right-handed (RH) neutrinos. Ref.~\cite{Accomando:2017qcs} studied the same processes $pp\to Z^\prime \rightarrow N \ N$ while aiming at prompt, short-lived RH neutrinos. Ref.~\cite{Batell:2016zod} studies the processes leading to displaced vertices at the LHC and demonstrates that the SHiP~\cite{Alekhin:2015byh, Anelli:2015pba} detector yields a competitive probe to constrain the active-sterile mixing. A displaced vertex search at LHCb has been proposed to probe boosted light heavy neutrinos~\cite{Antusch:2017hhu}. Ref.~\cite{Deppisch:2018eth} shows the potential to probe RH neutrinos with displaced vertices at the current LHC run, high luminosity run and proposed lepton colliders with a limit $V_{\mu N} < 10^{-7}$. The production of $B-L$ gauge boson $Z^{\prime}$ at lepton colliders has also been studied at Ref.~\cite{Ramirez-Sanchez:2016ugz}. Several studies have been focused on the seesaw aspect of the $B-L$ model. The local $B-L$
scenario is applied to a version of low scale seesaw mechanism at Ref. \cite{Khalil:2006yi}, an inverse seesaw scenario at Ref. \cite{Khalil:2010iu} and linear seesaw scenario at Ref. \cite{Dib:2014fua}. Similar studies have also been done for different aspect of the model and the heavy neutrinos in Refs.~\cite{Das:2017flq,Das:2017deo,Das:2018tbd,Jana:2018rdf} including displaced vertex signatures as well. 

In this work, we study the process of $pp \to Z^{\prime}\to NN$ leading to displaced vertices via decays of heavy neutrinos in the region compatible with current constraints on \mzprime and \gprime. It is important to note that the production cross section of this process is directly proportional to \gprime and inversely proportional to \mzprime. After taking a suitable \gprime allowed by current experimental searches, we show that the process has reasonable production cross section and the right-handed neutrinos can be long-lived when the active-sterile neutrino mixing is suppressed leading to displaced vertices. We then compute future sensitivities to such a signature based on the geometry of LHCb and CMS, as well as proposed outer detectors such as MATHUSLA, FASER and CODEX-b. In addition, in our paper we also include a discussion of a hypothetical, extended version of the MAPP detector, which we refer to as MAPP$^*$.

The paper is organized as follows, in Section~\ref{blreview}, we briefly review the $B-L$ model and the exotic decays of \zprime to right-handed neutrinos $N_{i}$. Next, we use the {\tt darkcast}~\cite{Ilten:2018crw} tool to recast the dark photons searches and obtain the current limits on the $B-L$ model, in section~\ref{sec:darkcast}. We introduce the trigger requirements and geometrical information for the detectors under consideration in section~\ref{sec:detectors}. We discuss our procedure for simulating the displaced vertices of heavy neutrino from \zprime boson and resulting sensitivities at the mentioned detectors in section~\ref{sec:simu}. Finally, we conclude in  section~\ref{conclu}.

\section{The $B-L$ gauge model}
\label{blreview}

\subsection{Model setup and particle spectrum}
In addition to the particle content of the SM, the $U(1)_{B-L}$ model contains an Abelian gauge field $B^\prime_\mu$, a SM singlet scalar field $\chi$ and three RH neutrinos $N_i$. The gauge group is $SU(3)_c\times SU(2)_L \times U(1)_Y \times U(1)_{B-L}$, where the scalar and fermionic fields $\chi$ and $N_i$ have $B-L$ charges $B-L = +2$ and $-1$, respectively. The scalar sector of the Lagrangian consists of
\begin{align}\label{Ls}
	{\cal L} \supset (D^{\mu}H)^\dagger(D_{\mu}H) 
	                       + (D^{\mu}\chi)^\dagger D_{\mu}\chi 
	                       - {\cal V}(H,\chi),
\end{align} 
where $H$ is the SM Higgs doublet and $V(H,\chi)$ is the scalar potential given by 
\begin{align}
\label{VHX}
	{\cal V}(H,\chi) = m^2 H^\dagger H + \mu^2 |\chi|^2 + \lambda_1 (H^\dagger H)^2 
	          + \lambda_2 |\chi|^4 + \lambda_3 H^\dagger H |\chi|^2.
\end{align}
Here, $D_\mu$ is the covariant derivative~\cite{Pruna:2011me}, 
\begin{align}
\label{DM}
	D_\mu = \partial_{\mu} + ig_{s}\mathcal{T}_\alpha G_\mu^\alpha 
	      + igT_a W_\mu^a + ig_1 Y B_\mu + i (\tilde{g}Y + g_1^{\prime} Y_{B-L}) B^\prime_\mu ,   
\end{align} 
where $G^\alpha_\mu$, $W^a_\mu$, $B_\mu$ are the usual SM gauge fields with associated couplings $g_s$, $g$, $g_1$ and generators $\mathcal{T}_\alpha$, $T_a$, $Y$. $B^\prime_\mu$ is the gauge field associated with the additional $U(1)_{B-L}$ symmetry with gauge strength $g_1^{\prime}$ and the $B-L$ quantum number $Y_{B-L}$. In this paper, we neglect the mixing between $U(1)_{B-L}$ and $U(1)_{Y}$ to simplify the model, i.e. we consider the minimal gauged $B-L$ model. This model is therefore valid in the limiting case of small mixing between SM photon and the $U(1)_{B-L}$ gauge boson. This crucially leads to  $Z^{\prime}$ production from quark-antiquark initial state via the $B-L$ charges. Consequently, the gauge sector of the model now includes the kinetic term  
\begin{align}
\label{LYM}
	{\cal L} \supset
	-\frac{1}{4} F^{\prime\mu\nu} F_{\mu\nu}^\prime,
\end{align} 
with the field strength tensor of the $B-L$ gauge group $F^\prime_{\mu\nu} = \partial_\mu B^\prime_\nu - \partial_\nu B^\prime_\mu$. This is manifest observationally as a new gauge boson, $Z^{\prime}$, coupling to SM fermions with a characteristic coupling $g_1^{\prime}$.
The fermion part of the Lagrangian now contains a term for the right-handed neutrinos
\begin{align}
\label{Lf}
	{\cal L} \supset
    i\overline{\nu_{Ri}}\gamma_\mu D^\mu \nu_{Ri}, 
\end{align} 
but is otherwise identical to the SM apart from the covariant derivatives incorporating the $B-L$ gauge field and the charges $Y_{B-L} = 1/3$ and $-1$ for the quark and lepton fields, respectively. Here, a summation over the fermion generations $i = 1, 2, 3$ is implied. Finally, the Lagrangian contains the additional Yukawa terms
\begin{align}
\label{LY}
	{\cal L} \supset 
	   - y_{ij}^\nu \overline{L_i}\nu_{Rj}\tilde{H}
	   - y_{ij}^M \overline{\nu^c_{Ri}} \nu_{Rj}\chi 
	   + \text{h.c.},
\end{align} 
where $L$ is the SM lepton doublet, $\tilde{H} = i\sigma^2 H^\ast$ and a summation over the generation indices $i,j = 1, 2, 3$ is implied again. The Yukawa matrices $y^\nu$ and $y^M$ are a priori arbitrary; the RH neutrino mass is generated due to breaking of the $B-L$ symmetry, with the mass matrix given by $M_R = y^M \langle\chi\rangle$. The light neutrinos mix with the RH neutrinos via the Dirac mass matrix $m_D = y^\nu v/\sqrt{2}$. The complete mass matrix in the $(\nu_L, \nu_R)$ basis is then
\begin{align}
\label{MD}
	{\cal M} = 
	\begin{pmatrix}
		0   & m_D \\
		m_D & M_R
\end{pmatrix},
\end{align} 
where 
\begin{align}
\label{MDM}
	m_D = \frac{y^\nu}{\sqrt{2}}v, \quad M_R = \sqrt{2} y^M x. 
\end{align} 
Here, $v = \langle H^0 \rangle$ and $x = \langle\chi\rangle$ are the vacuum expectation values for electroweak and $B-L$ symmetry breaking, respectively. In the seesaw limit, $M_R \gg m_D$, the light and heavy neutrino masses are 
\begin{align}
\label{seesaw}
	m_\nu \sim - m_D M^{-1}_R m^T_D, \quad m_N \sim M_R.
\end{align}
The flavour and mass eigenstates of the light and heavy neutrinos are connected as  
\begin{align}
\label{Neutrino}
	\begin{pmatrix}
		\nu_L \\ \nu_R
	\end{pmatrix} = 
	\begin{pmatrix}
		V_{LL} & V_{LR} \\
		V_{RL} & V_{RR}
	\end{pmatrix}
	\begin{pmatrix}
		\nu \\ N
	\end{pmatrix},
\end{align} 
schematically writing the 6-dimensional vectors and matrix in terms of 3-dimensional blocks in generation space. The mixing and the light neutrino masses are constrained by oscillation experiments to yield their observed values, i.e. the SM charged current lepton mixing $V_{LL} = U_\text{PMNS}$ (apart from small non-unitarity corrections and assuming the charged lepton mass matrix to be diagonal). For the case of one generation of a light and heavy neutrino we will consider in turn, this reduces to the $2\times 2$ matrix form 
\begin{align}\label{Rotation}
	\begin{pmatrix}
		\nu_L \\ \nu_R
	\end{pmatrix} = 
	\begin{pmatrix}
    	\cos\theta_\nu & -\sin\theta_\nu \\
		\sin\theta_\nu &  \cos\theta_\nu
	\end{pmatrix}
	\begin{pmatrix}
		\nu \\ N
	\end{pmatrix}.
\end{align}
For simplicity, we thus neglect mixing among flavours and therefore generations decouple. The Yukawa coupling matrix then becomes diagonal and we can write ($i = e, \mu, \tau$)
\begin{align}
\label{YY}
	y^\nu_{ii} = \frac{\sqrt{2} m_{N_i} V_{iN}}{v},
\end{align} 
using the neutrino seesaw relation. Here, $V_{iN}$ represents the active-sterile mixing, $\sin\theta_i = V_{iN}$, in the three generations, $V_{e N}$, $V_{\mu N}$, $V_{\tau N}$. 

Similar to the light and heavy neutrinos, the additional scalar singlet $\chi$ also mixes with the SM Higgs. The mass matrix of the Higgs fields $(H, \chi)$ at tree level is~\cite{Robens:2015gla}
\begin{align}
\label{mass}
	M_h^2 = \begin{pmatrix}
		2\lambda_1 v^2 & \lambda_3 x v \\
		\lambda_3 x v & 2\lambda_2 x^2
	\end{pmatrix}.
\end{align}
The physical masses of the two Higgs $h_1, h_2$ are then 
\begin{align}
\label{Higgsmass}
	m^2_{h_{1(2)}} = \lambda_1 v^2 + \lambda_2 x^2 
			     -(+) \sqrt{(\lambda_1 v^2 - \lambda_2 x^2)^2 + (\lambda_3 xv)^2},
\end{align} 
and the physical Higgs states ($h_1,h_2)$ are related to the gauge states ($H, \chi$) as
\begin{align}
\label{Higgs mixing}
	\begin{pmatrix}
		h_1 \\ h_2
	\end{pmatrix} = 
	\begin{pmatrix}
    	\cos\alpha & -\sin\alpha \\
		\sin\alpha &  \cos\alpha
	\end{pmatrix}
	\begin{pmatrix}
		H \\ \chi
	\end{pmatrix}.
\end{align} 
The directly measurable parameters for the Higgs sector are the masses $m_{h_1}$ and $m_{h_2}$, as well as the mixing angle $\alpha$,
\begin{align}
\label{lambda}
	\tan(2\alpha) = \frac{\lambda_3 v x}{\lambda_2 x^2 - \lambda_1 v^2}.
\end{align}
%
\subsection{Production and decay of heavy neutrinos}
\label{sec:prodN}
We intend to analyze the displaced decay of heavy neutrinos produced via a $Z'$ boson, $pp \to Z' \to N N$. Both $Z'$ and $N$ have very small widths in our scenario and we can thus employ the narrow width approximation for the process cross section,
\begin{align}
	\sigma(pp \to Z' \to NN \to \text{DV}) = \sigma(pp \to Z') 
	\times \text{BR}(Z' \to N N)
	\times \text{BR}(N \to \text{visible}).
\end{align} 
Here, the observed final state consists of a single displaced vertex produced by one of the two heavy neutrinos decaying into a visible SM particle state, i.e. one which does not contain only light neutrino. For some of the detectors discussed later we will employ a more specific final state for the which we will use the corresponding branching ratio of $N$.

The partial decay width for $Z^{\prime}\to N_i N_i$ per generation of heavy neutrinos is
\begin{align}
\label{partialdecay}
	\Gamma(Z^{\prime}\to N N) = \frac{1}{6}\frac{(g_1^{\prime})^{2}}{4\pi} m_{Z^{\prime}}\left(1-\frac{4 m_N^2}{m_{Z^{\prime}}^2}\right)^{3/2},
\end{align} 
while for each SM fermion generation, assuming them to be massless, it is
\begin{align}
\label{partialdecay}
	 \Gamma(Z^{\prime}\to l_i\bar{l_i}) = 
	2\Gamma(Z^{\prime}\to \nu_i\bar{\nu_i}) = 
	3\Gamma(Z^{\prime}\to q_i\bar{q_i}) =
	\frac{1}{3}\frac{(g_1^{\prime})^{2}}{4\pi}m_{Z^{\prime}}
	\left(1-\frac{4 m_X^2}{m_{Z^{\prime}}^2}\right)^{3/2}.
\end{align} 
Thus the total decay width of $Z^{\prime}$ can be expressed as
\begin{align}
\label{totaldecay}
	\Gamma(Z^{\prime})\propto \bigg(2\times3+1\times3+(1/3)\times2\times5+3\times\Big(1-\frac{4 m_N^2}{m_{Z^{\prime}}^2}\Big)^{3/2}\bigg)\Gamma(Z^{\prime}\to \nu \nu)
\end{align}
Here, the factor $2\times3$ represents 3 generations of leptons while $1\times3$ corresponds to 3 generations of light Majorana neutrinos. The $B-L$ charge of quarks is 1/3, and for $m_{Z^{\prime}}$ relatively small, only five possible flavours of quarks are accessible.

However, these analytical calculations have flaws such that it does not consider the effects of hadronization while the quarks pair final states can consist into hadrons. These effects can be better described in Vector Meson Dominance (VMD) mechanism~\cite{Fujiwara:1984mp}. Later in this paper, we will be directly using the branching ratio of $Z^{\prime}$ boson decays in a tool called {\tt Darkcast}~\cite{Ilten:2018crw} which counts the effects of hadronization using VMD while heavy neutrinos final states are manually added as invisible decays into the code.

For the model presented here, we have four parameters of interest, the mass of \zprime, $m_{Z^\prime}$, the mass of heavy neutrino $m_N$, the $U(1)_{B-L}$ gauge coupling $g_1^\prime$ and the active-sterile mixing $V_{i N}$. We assume three generations of heavy neutrinos each mixed with the corresponding leptons such that only decays into the SM final states are allowed through active-sterile mixing $V_{iN}$. Such a setup also forbids mixing between generations, leading to a block diagonal mixing matrix between active-sterile neutrinos in generation space. In order to reduce the degrees of freedom in our analysis, we assume, $\frac{m_N}{m_{Z^{\prime}}}=0.3$. This choice opens the $Z^{\prime}\rightarrow NN$ decay channel. Furthermore we assume that there are no other light exotic particles, such that no modification of the $Z^\prime \to NN$ decay rates is possible. With the assumption of flavour-diagonal active-sterile neutrino mixing, we concentrate on one generation of heavy neutrinos, namely the muon neutrinos. Therefore, the decay modes of our interest are three body processes such as $N\to \mu^\pm q\bar{q}$ and $N\to \mu^+\mu^-\nu_\mu$ via  $W^{\pm(*)}, Z^{(*)}$. The branching ratio of heavy neutrinos have been already shown in the Ref.~\cite{Deppisch:2018eth}. Approximately, the resulting decay length for $m_N \lesssim m_Z$ can be expressed as 
\begin{align}
\label{lengthapproxi}
	L_N \approx 0.025~\text{m} 
	\cdot \left(\frac{10^{-6}}{V_{\mu N}}\right)^2 
	\cdot \left(\frac{100~\text{GeV}}{m_N}\right)^5,
\end{align} 
As an example, for a RH neutrino mass $m_N \approx 3$~GeV and a mixing $V_{\mu N} \approx 10^{-2}$, the decay length is of the order $L_N \approx 0.1$~mm. For smaller mixing, such as in the naive seesaw estimate $V_{\mu N} \approx 10^{-4}$, the decay length can be of the order of meters, potentially detectable through a displaced vertex at the proposed far detectors. For even smaller mixing and thus longer decay lengths, the decay products will escape the detector volume resulting in a missing energy signature.

Despite having a long decay length, the potential of the proposed and current detectors to probe such long lived scenarios depends on two things: (i) the production cross section of the heavy neutrinos given by the product of the \zprime production cross section and the branching ratio $BR(Z^{\prime} \to N N)$ and (ii) the geometry of the detector. While the importance of the production cross section is clear, the dependence on detector geometry is less straightforward. Primarily, in this scenario, we have two mass scales, the \zprime mass $m_{Z^\prime}$ and the mass of heavy neutrino $m_N$. The final state products of importance to us are the decay products of the heavy neutrinos, i.e. muons. While the displacement of the muons depends on the lifetime of the heavy neutrino, the boost of these muons depends on the mass hierarchy between \mzprime, $m_N$ and the mass of muon. This is illustrated in the Fig. \ref{fig:distro}.

Before we proceed with a study of several lifetime frontier detectors, we first estimate the signal strengths and kinematics of the events. To this extent, we look at the production cross section of the \zprime\,, the decay of \zprime\, and heavy neutrino ($N$), and a few select kinematic variables to illustrate the effect of mass hierarchies within the scenario under consideration. 

\begin{figure}
\centering
\includegraphics[width=0.49\textwidth]{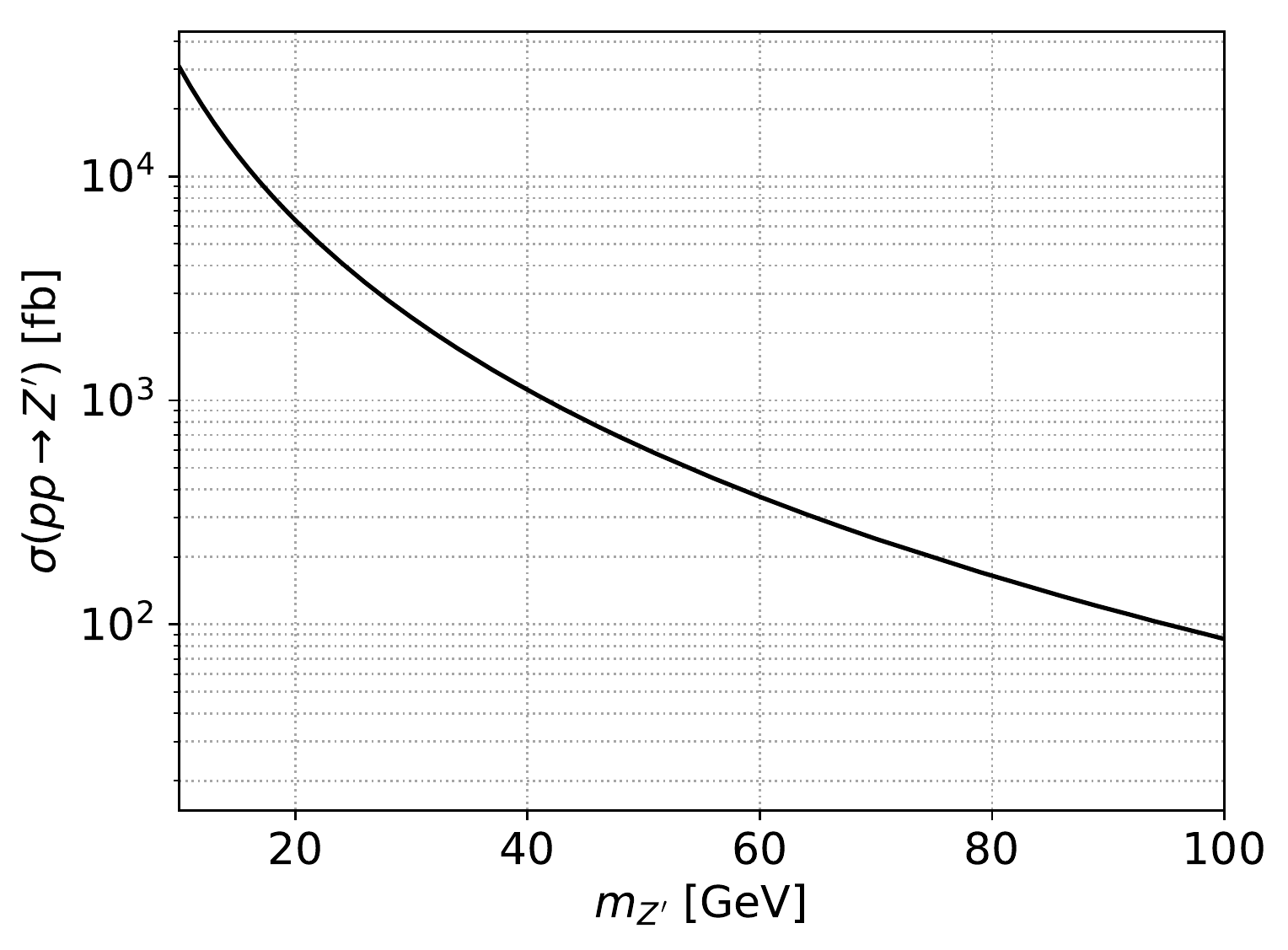}
\includegraphics[width=0.49\textwidth]{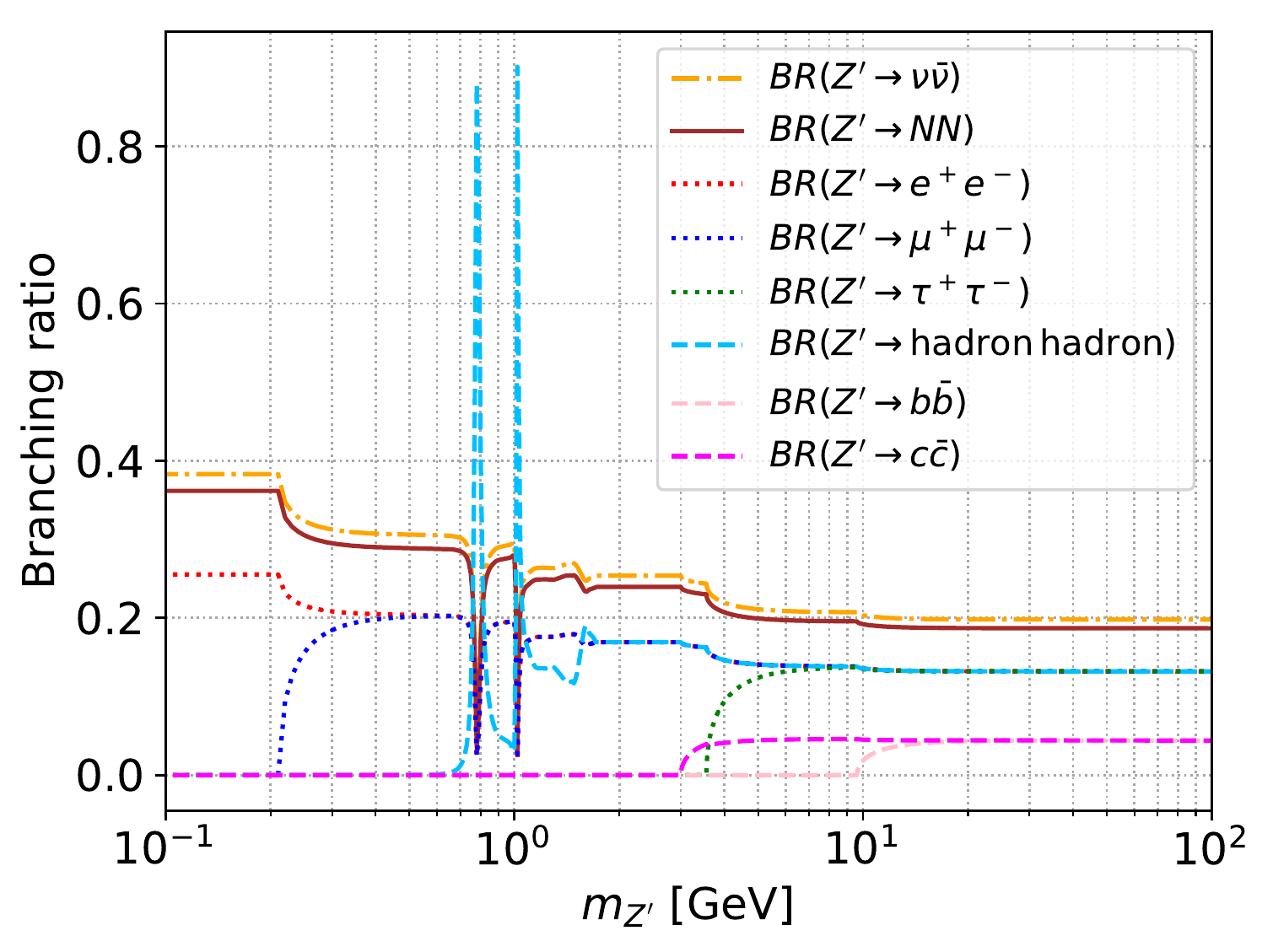}
\caption{Production cross section of $pp \to Z'$ at $\sqrt{s} = 14$~TeV for \gprime  = $10^{-3}$ (left). Branching ratios of \zprime to various final states; here $\nu\bar\nu$ and $NN$ represent the sum over all three light and heavy neutrinos in the corresponding final states (right).}
\label{fig:blxsecbr}
\end{figure}
Figure~\ref{fig:blxsecbr}, shows the production cross section and the branching ratios of the \zprime\, gauge boson for a wide range of masses. The production cross section (fig. ~\ref{fig:blxsecbr}, left) varies from a few pb for very light \zprime\, to $\mathcal{O}(100)$ fb for heavy \zprime for a \gprime coupling of $10^{-3}$. We also plot the branching ratios of \zprime\, to several SM and exotic final states in the right panel. Here three degenerate heavy neutrino species have been assumed. The assumption of degenerate neutrino species does not hold throughout this paper. In case only one light heavy neutrino species is assumed, the corresponding \zprime\, branching ratio is about 8\%. These numbers demonstrate that it is viable to reach displaced heavy neutrino decays during HL-LHC program.

\subsection{Constraints on the $B-L$ model}
\label{sec:darkcast}
Before we proceed to demonstrate the potential of future detectors, it is imperative to look at the existing constraints on the $m_{Z^\prime}$ - \gprime parameter space. We use the publicly available reinterpretation tool {\tt Darkcast} in order to derive these constraints. Initial demonstration of the capabilities of {\tt Darkcast} contains reinterpretation to $B-L$ model without heavy neutrinos. Adding heavy neutrino $N$ as final state in $Z^{\prime}$ decays can be achieved by adding invisible decay $\Gamma_{Z^{\prime}\to N \ N} = 3\times 0.944\times\Gamma_{Z^\prime \to \nu_e \ \nu_e}$ where the 0.944 is the phase space factor assuming $\frac{m_N}{m_{Z^{\prime}}} = 0.3$ to the $B-L$ boson model. Unlike focusing on the second generation of heavy neutrino elsewhere in this paper, we take 3 generations here because all of three generations would reduce the visible branching ratio of $Z^{\prime}$ thus affect the limits in the {\tt Darkcast}. As the most stringent limits in {\tt Darkcast} are from prompt searches, we assume the heavy neutrinos to be stable at the scale. While the original $B-L$ model in the {\tt Darkcast} is applicable to $B-L$ parameters space regions where the heavy neutrino masses are bigger than the $Z^{\prime}$ bosons thus $Z^{\prime}$ decays to heavy neutrinos are kinematically forbidden.

\begin{figure}
	\centering
	\includegraphics[width=0.49\textwidth]{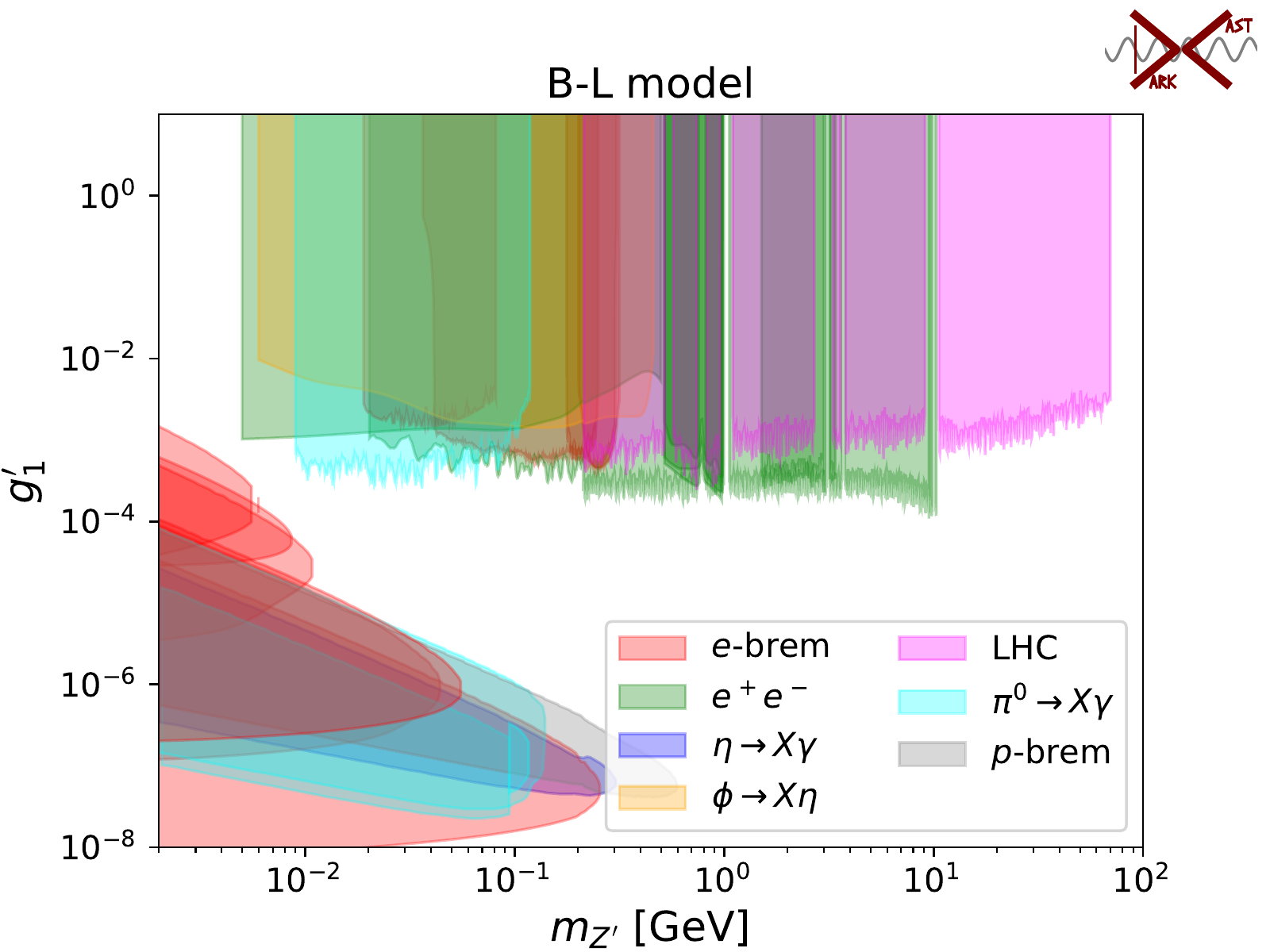}
	\includegraphics[width=0.49\textwidth]{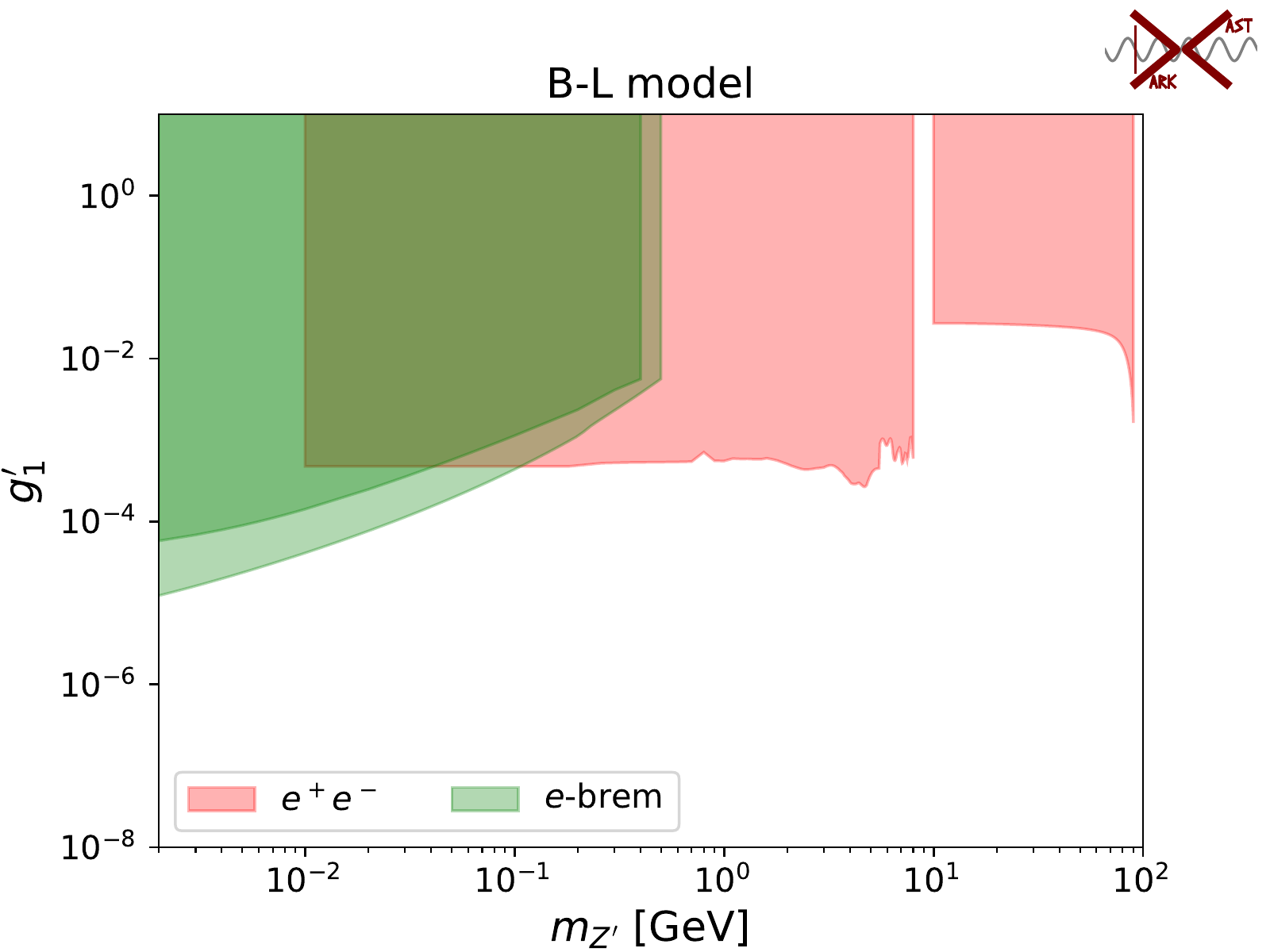}
	\caption{Existing limits on the \gprime - $m_{Z^{\prime}}$ parameter space using {\tt DarkCast}. Constraints derived from decays to SM final states for a $B-L$ model containing three generations of heavy neutrinos (left). Constraints derived from $Z'$ decays to heavy neutrino final states assuming that the heavy neutrinos are stable.}
	\label{fig:userzp}
\end{figure}
Defining heavy neutrino final states as invisible signals, we can evaluate the impact of existing searches on the \gprime - $m_{Z^{\prime}}$\, parameter space \ref{fig:userzp}. We show that the limits on $B-L$ model from the {\tt Darkcast} do not change drastically as the branching ratios differ by only $\sim$ 10$\%$. We therefore use following limits on the \gprime - $m_{Z^{\prime}}$\, parameter space throughout the rest of this paper as:
\begin{itemize}
	\item $g_1^{\prime} \leq 10^{-3}$ for $m_{Z^{\prime}}> 10$ GeV
	\item $g_1^{\prime} \leq 10^{-4}$ for $m_{Z^{\prime}}< 10$ GeV 
\end{itemize}
With these limits from {\tt Darkcast} we choose a set of benchmark corresponding to $m_{Z^{\prime}} - g_1^{\prime}$ as $g_1^{\prime}$ = $10^{-3}$ for $m_{Z^{\prime}}> 10$ GeV and $g_1^{\prime}$ = $10^{-4}$ for $m_{Z^{\prime}}< 10$ GeV. This fixes two of the four free parameters of our model. While these limits already put some constraints on the parameter space of our interest, the question of assessing impact of completely stable heavy neutrino remains. It is useful to do this recasting nonetheless as it demonstrates that neglecting the \zprime\, decays to heavy neutrino has little impact on the parameter space. Finally, such a recast leads to a benchmark set of \gprime - $m_{Z^{\prime}}$\, which will be used from here on. 

Going forward, we will give up on the assumption of stable heavy neutrinos and assess the sensitivity of future lifetime frontier detectors in for the long lived heavy neutrinos. In order to do that, we first illustrate our analysis setup in the next section.

\section{Analysis setup}
\label{sec:detectors}
\subsection{Signal generation}
We take the benchmark parameter space of $g_1^{\prime}$ and $m_{Z^{\prime}}$ in the $B-L$ model, as $g_1^{\prime}$ = $10^{-4}$ for $m_{Z^{\prime}}< 10$ GeV and $g_1^{\prime}$ = $10^{-3}$ for $m_{Z^{\prime}}> 10$ GeV. Unlike the searches targeting prompt final states of this $Z^{\prime}$ decays, the decays leading to displaced final states, either when $Z^{\prime}$ is displaced or the heavy neutrino is displaced, have low to no backgrounds. In this section, we consider such displaced decays at all the detectors mentioned in the last section. Due to the expected lack of factorization, the Monte Carlo simulation for such low mass resonances below 10~GeV is likely to involve non-perturbative processes. Thus, we only simulate the case where $m_{Z^\prime} > 10$~GeV.

We use the Universal FeynRules Output (UFO)~\cite{Degrande:2011ua} of $B-L$ model developed in the Ref.~\cite{Deppisch:2018eth} in combination with the Monte Carlo event generator {\tt MadGraph5aMC$@$NLO} -v2.6.3~\cite{Alwall:2014hca} at parton level. For every signal sample, we generate 100k signal events. We then pass the generated parton level events on to PYTHIA v8.235~\cite{Sjostrand:2014zea} which handles the initial and final state parton shower, hadronization, heavy hadron decays, etc. It should be noted that no detector simulation is taken into account. It is clear that these assumptions are idealistic, however, in order to assess the reach and demonstrate the complementarity of different lifetime frontier experiments, such an analysis should be sufficient. We produce three different processes $pp \to Z^{\prime}\to NN$ and decay only one of the heavy neutrino as either $N\to \mu^\pm q\bar{q}$ for LHCb or $N\to \mu^+\mu^-\nu_\mu$ for CMS, while we do not decay the neutrinos at all for estimating sensitivity at CODEX-b, MAPP$^{*}$ and MATHUSLA. For CODEX-b, MAPP$^{*}$ and MATHUSLA we assume that every decay of heavy neutrinos except the fully invisible mode can be detected. It will be computationally expensive to generate every point in the mass and mixing parameter space under consideration. In order to simplify our signal generation, we generate signal samples with prompt decays of neutrinos and after we compute the vertex information given the rest lifetime of the heavy neutrinos  and momentum of particles as given in the HepMC file. To simulate the exponential decay of heavy neutrinos, we generate $100 \times \sigma \times \mathcal{L} / 10^4$ number of distributions of exponential decay for each heavy neutrino event. The events are passed through the kinematic and geometric selection criteria corresponding to the different detectors. These are described below. For each of the detectors, any neutrino decaying within the specified volume is assumed to be detected.
\subsection{Displaced vertex detectors}
\label{sec:LHCb}

\paragraph{CMS}
ATLAS and CMS are two general purpose detectors at the LHC. They are both capable of probing displaced vertex signatures up to moderate displacements. Here we take CMS for simplicity. The CMS analysis we considered requires two muons in the final state, correspondingly we consider the fully leptonic decay of the heavy neutrino. The kinematical cuts can be summarised as two muon tracks that satisfy the following constraints on the transverse momentum of the leading ($\mu_1$) and sub-leading muon ($\mu_2$), pseudo-rapidity $\eta$ and isolation $\Delta R$ of the two tracks,~\cite{Deppisch:2018eth, Accomando:2016rpc,CMS:2015pca,CMS:2014hka}:
\begin{gather}
	p_T(\mu_1) > 26\text{ GeV}, ~~
	p_T(\mu_2) > 5\text{ GeV}, ~~
	|\eta| < 2.0 ~~
	\Delta R > 0.2,~~
	\cos\theta_{\mu\mu} > -0.75.
	\label{cutsLHC}
\end{gather}
Here the cut on $\Delta R$ ensures, that the detected muon  corresponds to the muon identified by the trigger. Background due to cosmic ray muons can be rejected efficiently by correlating the corresponding hits with the beam collision time and with the cut on the angle between the muons, $\cos\theta_{\mu\mu}$ \cite{CMS:2015pca}. An additional cut on the difference $\Delta\Phi < \frac{\pi}{2}$ in the azimuthal angle between the dilepton momentum vector and the vector from the primary vertex to the dilepton vertex has also been applied in the Refs.~\cite{CMS:2014hka, CMS:2015pca}, which we do not employ here as it does not affect our analysis.

In addition to the kinematical cuts, we also implement geometric cuts in order to simulate the geometric coverage of the detector. We use the following characteristics for Region~1 and 2 to represent a typical LHC detector~\cite{Accomando:2016rpc}, where the variables used are discussed above,
\begin{align}
\label{cut1}
	\text{Region 1:}&\quad
	10~\text{cm} < |L_{xy}| < 50~\text{cm},\ |L_z| < 1.4~\text{m},\
	d_0/\sigma_d^t > 12,\ \sigma_d^t = 20~\mu\text{m}, \\
\label{cut2}
	\text{Region 2:}&\quad
	0.5~\text{m} < |L_{xy}| < 5~\text{m},\ |L_z| < 8~\text{m},\ 
	d_0/\sigma_d^t > 4,\ \sigma_d^t = 2~\text{cm}.
\end{align}
Here the transverse distance $|d_0| = |x p_y - y p_x|/p_T$, where $x$ and $y$ are the position where the right handed neutrino decayed, and $p_x$, $p_y$, $p_T$ are the components of momentum and transverse momentum of the final particles $\mu$, and $L_{xy}$ / $L_z$ are the transverse / longitudinal decay lengths of the RH neutrino, and $\sigma_d^t$ is the resolution of the detector in transverse distance. 

\paragraph{LHCb}
The displaced signatures of the heavy neutrinos $N$ from $Z^{\prime}$ at the LHCb can be simulated while using a similar trigger requirement of the current LHCb displaced vertices searches~\cite{Antusch:2017hhu, Aaij:2016xmb}: $N(\mu) = 1$ and $N(j) > 0$, $p_T(\mu) > 12$~GeV, $M[\mu jj] > 4.5$~GeV and $2<\eta(\mu,jets) < 5$.

The geometrical selections corresponding the LHCb detector are set as~\cite{Antusch:2017hhu}:
\begin{itemize}
	\item arctan $\frac{L_{xy}}{L_z}$ < 0.34 for the direction of the displacement.
	\item Region 1 restricts the decays to be inside the VELO (VErtex LOcator), for which we take 0.02 m < $r$ < 0.5 m, $L_z$ < 0.4 m, where $r$ is the radius. Here we assume a signal reconstruction efficiency of 100$\%$.
	\item  Region 2 restricts the decays to be inside the volume with 0.005 m < $r$ < 0.6 m and $L_z$ < 2 m, which is the radial extension and distance to the TT tracking station. For the region which is not already included in Region 1, due to, e.g., detector related backgrounds and blind spots, we use a 50$\%$ signal reconstruction efficiency. 
\end{itemize}

\paragraph{MATHUSLA}
There are several proposals to equip the high luminosity run of the LHC (HL-LHC) with additional detectors to search for long-lived particles. We consider multiple such options.  One such example is the proposal comprising of a large detector on the ground surface called MATHUSLA (MAssive Timing Hodoscope for Ultra Stable neutraL pArticles), to detect ultra long-lived particles a few hundred meters away from the collision point~\cite{Chou:2016lxi}. We estimate the sensitivity of this setup by applying one the following geometrical selection cuts: 
\begin{gather}
	-100~\text{m} < L_x < 100~\text{m},\, 
	 100~\text{m} < L_y < 120~\text{m},\, 
	 100~\text{m} < L_z < 300~\text{m}.
\end{gather} 
Due to its setup, MATHUSLA has a comparatively small geometric coverage. However, it can potentially probe very small active-sterile neutrino mixing as it would be situated far away from the interaction point.

\paragraph{FASER}
The FASER (the ForwArd Search ExpeRiment) detector is placed 480 meters from the Interactive Points (IP) centred at the beam direction. The geometry of its two phase design are proposed as~\cite{Ariga:2018uku}:
\begin{itemize}
	\item FASER 1: A cylinder with 1.5 m of its depth at the beam direction with a 0.1 m radius running at 150 fb$^{-1}$ LHC. 
	\item FASER 2: A cylinder with 5 m of its depth at the beam direction with a 1 m radius running at 3000 fb$^{-1}$ HL-LHC.
\end{itemize}
The FASER 1 phase will not have sufficient sensitivity towards heavy neutrinos in our model, and we will concentrate on FASER 2 in our simulations.

\paragraph{MAPP$^{*}$}
The MAPP (Monopole Apparatus for Penetrating Particles) detector is a planned sub-detector of the MoEDAL experimental setup~\cite{Pinfold:2019nqj}. It is to be operated at a 300~fb$^{-1}$ and 14~TeV HL-LHC placed about 50~meters from the LHCb interaction point with the depth of the tunnel ranging from 7 to 10 meters at an oblique angle to the beam-line ranging from 5 to 25 degrees between the beam direction. The original design of the detector is consists of two arrays of scintillator bars (with detector volume of 1m$^3$ ). This detector can move inside the whole tunnel.

With the setup described above, MAPP will not have sufficient sensitivity toward heavy neutrinos in our model. We instead simulate the sensitivity for a hypothetical version of the MAPP detector covering the whole available tunnel. We refer to this detector option as MAPP$^{*}$.

\paragraph{CODEX-b}

CODEX-b (the COmpact Detector for EXotics at LHCb) has been proposed to be constructed in a large unoccupied space after the upcoming Run 3 upgrade of LHCb, shielded by a 3 m thick concrete radiation shield~\cite{Gligorov:2017nwh}. In order to detect displaced vertices, we require that the heavy neutrinos to decay inside the detector volume~\cite{Berlin:2018jbm}: 
\begin{gather}
	26~\text{m} < L_x < 36~\text{m},\, 
	 -3~\text{m} < L_y < 7~\text{m},\, 
	 5~\text{m} < L_z < 15~\text{m}.
\end{gather} 
We note that Refs.~\cite{Berlin:2018jbm, Gligorov:2017nwh} use an additional cut on the track energy to be above a threshold of 600~MeV. This cut does not have any impact for the relatively large masses in our case.

\section{Results}
\label{sec:simu}
\begin{figure}[t!]
\centering
\includegraphics[width=0.49\linewidth]{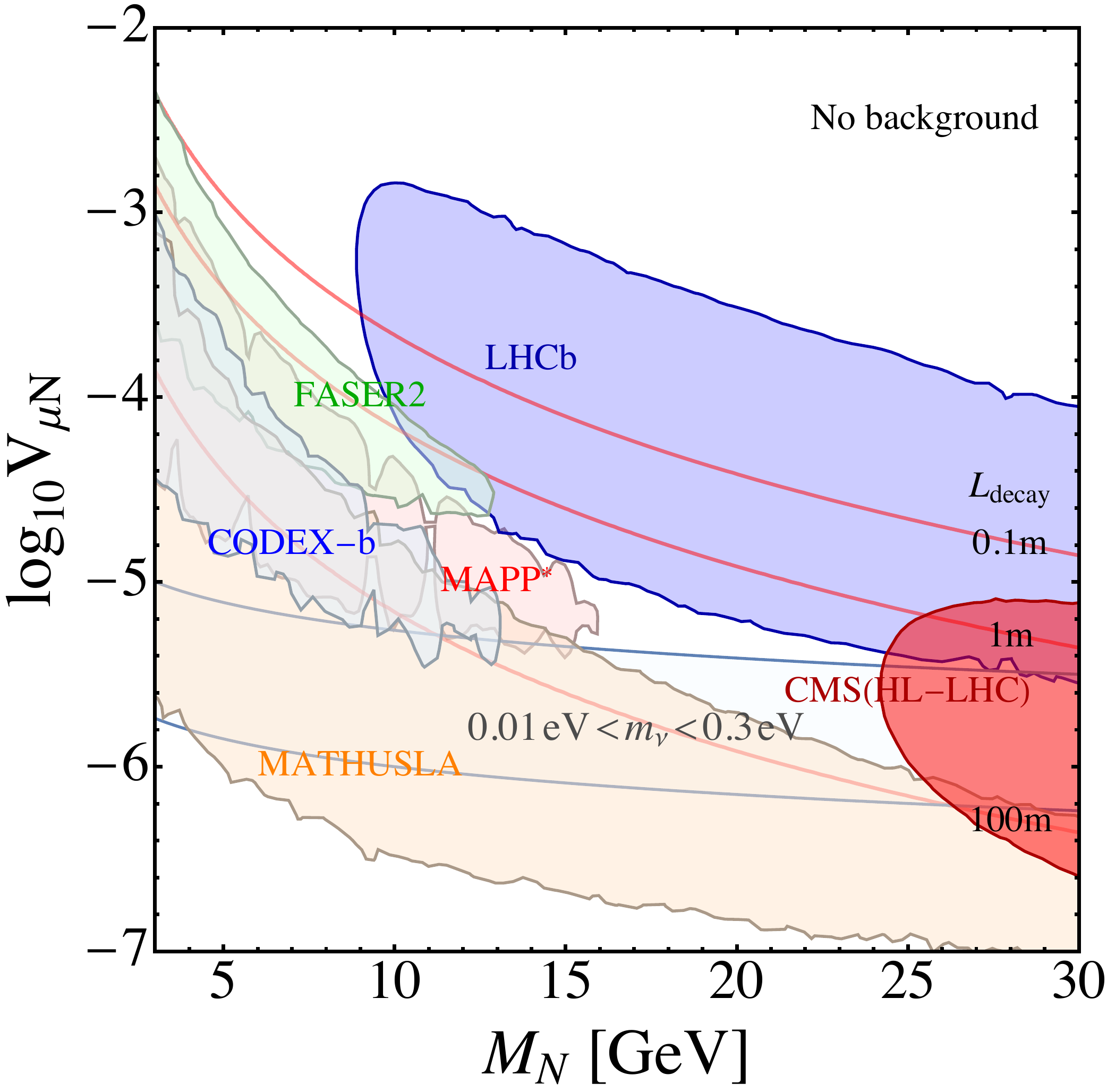}
\includegraphics[width=0.49\linewidth]{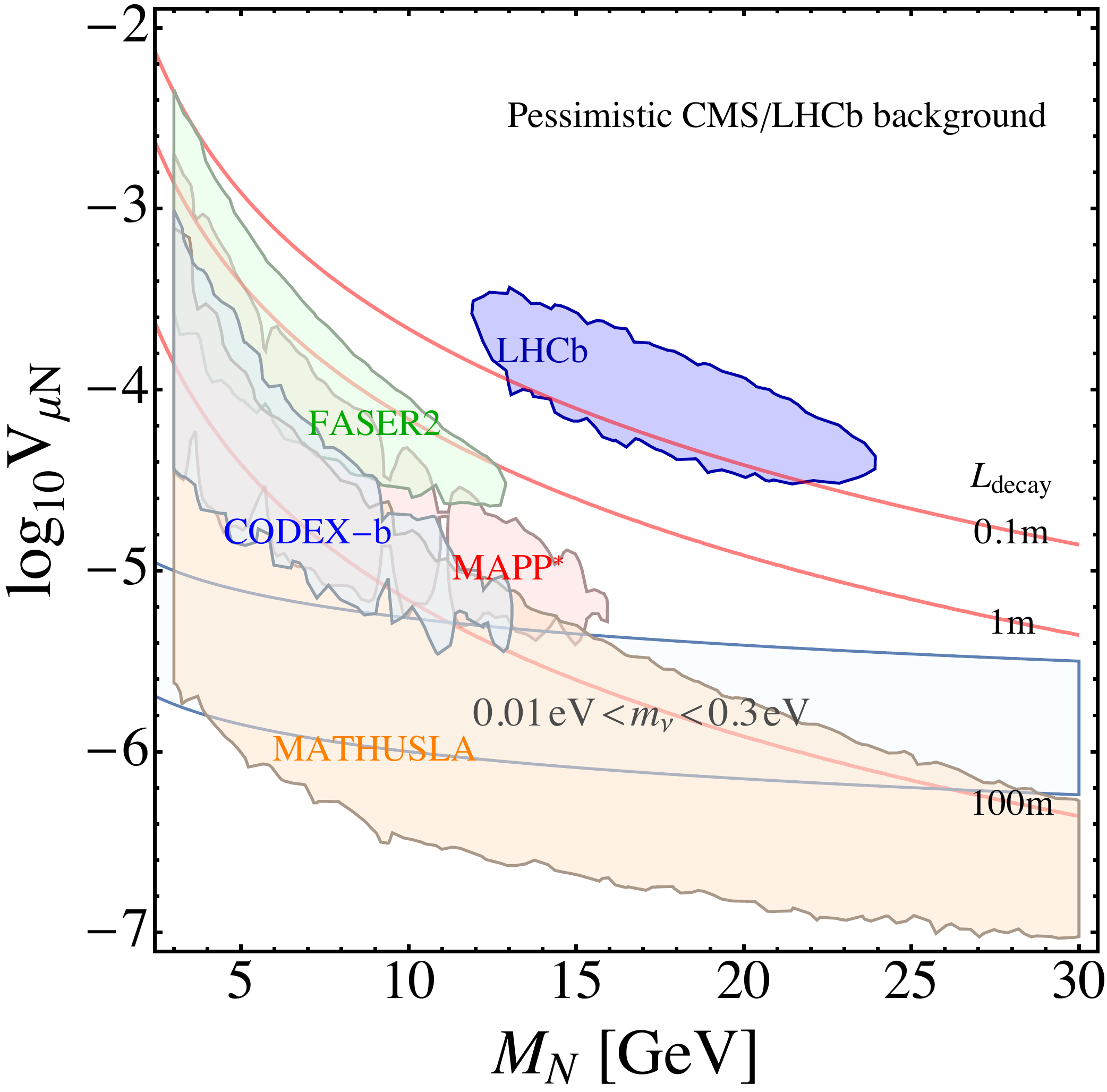}
\caption{Projected sensitivity of displaced vertex detectors in the heavy neutrino mass $m_N$ and neutrino mixing $V_{\mu N}$ parameter plane. The $U(1)_{B-L}$ gauge coupling and the $Z'$ mass are chosen as $g_1^{\prime} = 10^{-3}$ and $m_{Z^{\prime}} = 3.33\times m_N$, respectively. The displaced vertex detectors are CMS (HL-LHC), LHCb, MATHUSLA, FASER 2, MAPP$^*$ and CODEX-b as indicated, at 14~TeV and using projected future luminosity as detailed on the text. In the left plot, we assume no background to displaced vertex searches whereas in the right plot, we scale current background limits for CMS and LHCb to the luminosity used. The red curves denote the proper decay length of the heavy neutrino $N$ and the horizontal band indicates the preferred parameter region where the light neutrinos acquire a mass between $10^{-2}$~eV and $0.3$~eV in a canonical seesaw mechanism. }
\label{fig:big}
\end{figure}
We now estimate the sensitivity of the above displaced detectors in probing the active-sterile neutrino mixing. We take 3000~fb$^{-1}$ luminosities for CMS and MATHUSLA, and 300~fb$^{-1}$ luminosity for LHCb, MAPP$^{*}$ and CODEX-b. For FASER, there are two design phases with 150~fb$^{-1}$ for FASER and 3~ab$^{-1}$ for FASER 2. We only consider the latter of the two.

\begin{figure}[t!]
\centering
\includegraphics[width=0.48\linewidth]{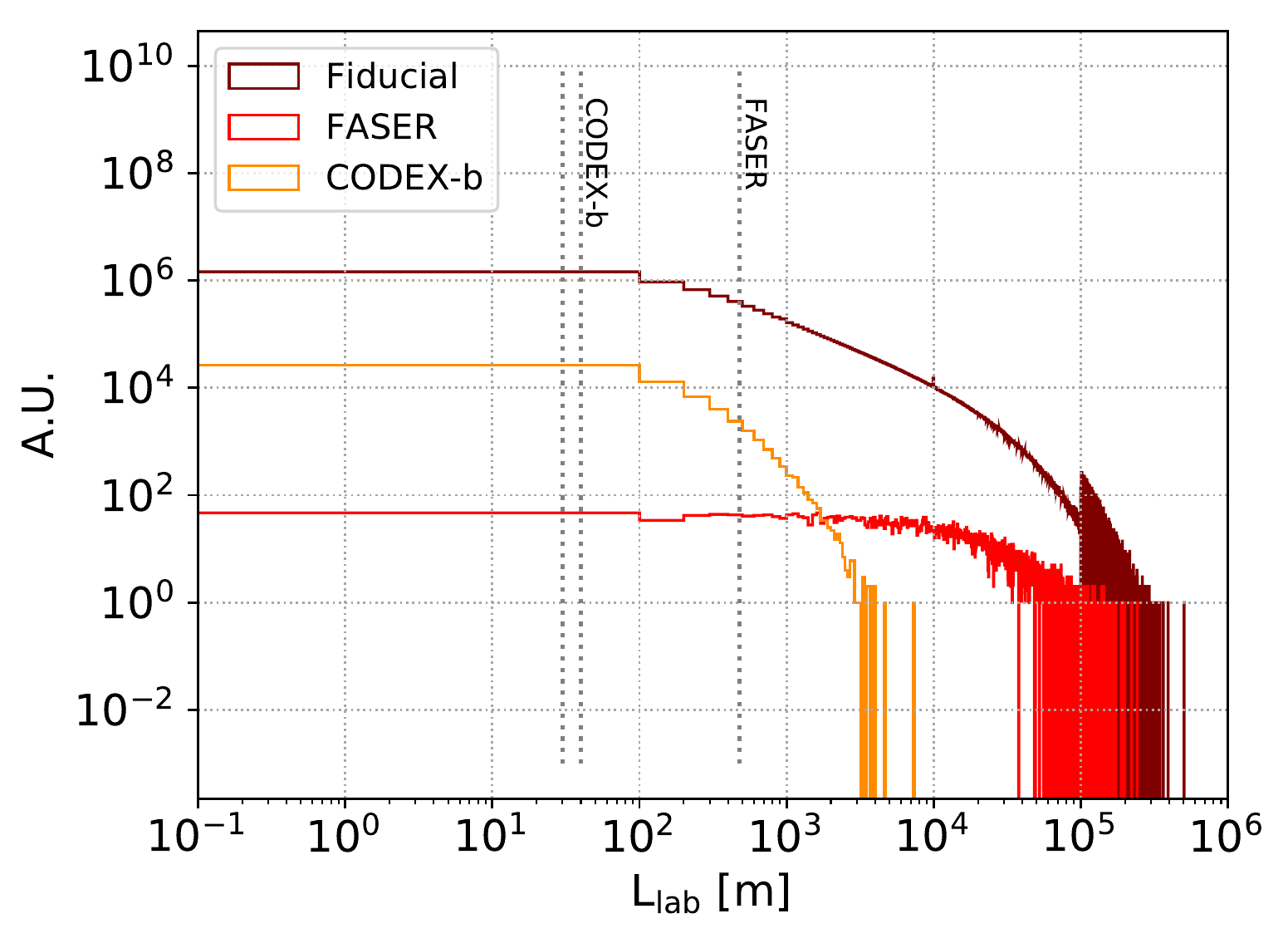}
\includegraphics[width=0.48\linewidth]{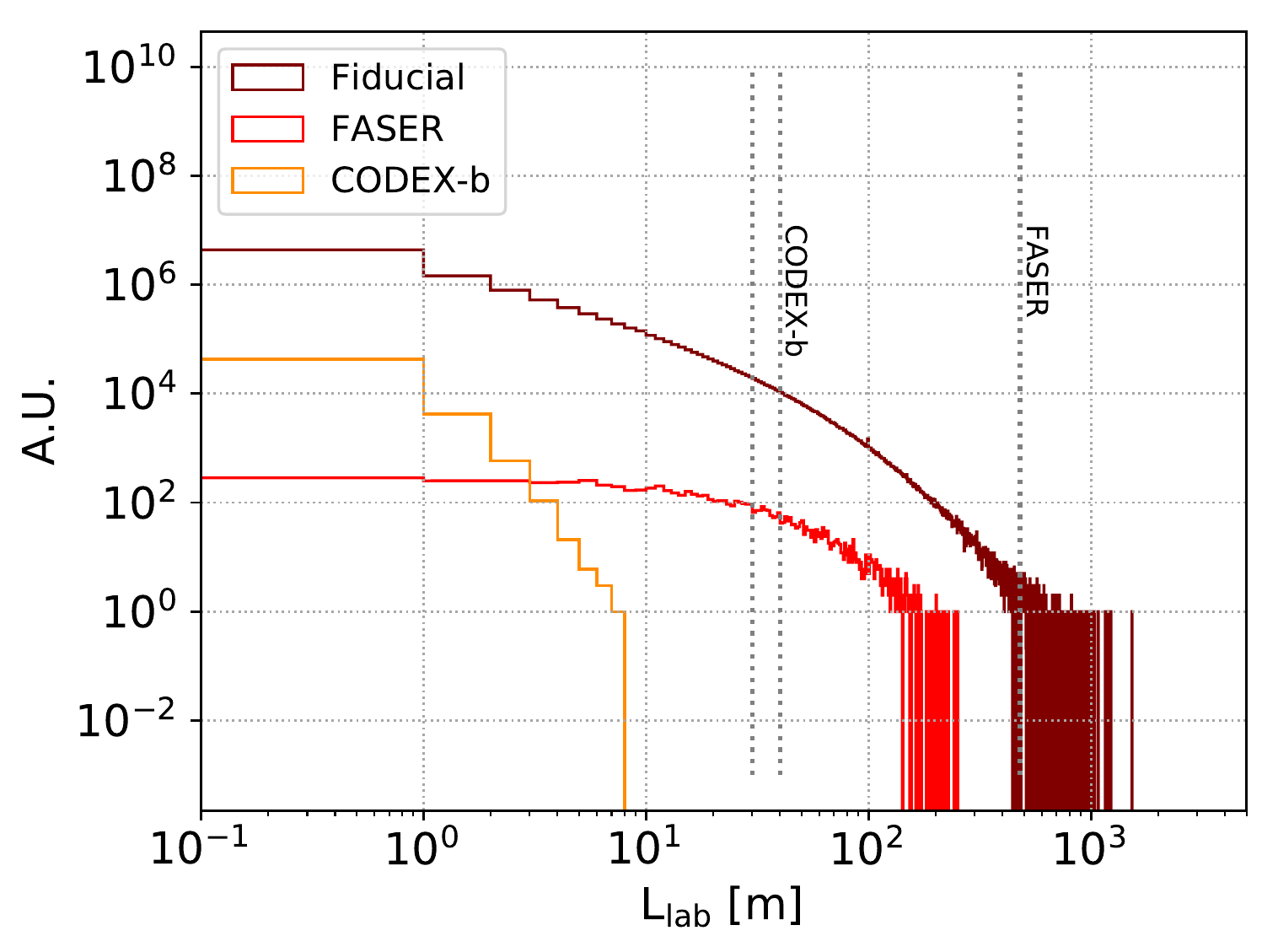}
\caption{Distribution of the heavy neutrino decay length in the laboratory system for  $m_N = 5$~GeV, $m_{Z^\prime} = 17$~GeV (left) and $m_N = 15$~GeV, $m_{Z^\prime} = 50$~GeV (right). The active-sterile neutrino mixing is set at $V_{\mu N} = 10^{-4.5}$. The three histograms represent the distributions of all events (fiducial) as well as those passing the angular acceptance criteria of FASER and CODEX-b. The vertical lines show the approximate distance from the interaction point at which FASER and CODEX-b are located. For CODEX-b, the two lines shown represents the depth of detector.}
\label{fig:histos}
\end{figure}

We first take an optimistic view assuming no backgrounds for the displaced vertex signatures. For detectors located far away from the IP, e.g. MATHUSLA, FASER, MAPP$^{*}$ and CODEX-b, this is quite safe due to the large displacement from the IP. Following a Poisson distribution, we take the parameter space which has the number of signal events above 3 (3.09) to be excluded~\cite{Agashe:2014kda} at 95 $\%$ C.L.. In the left panel of Figure~\ref{fig:big}, we show the resulting reach of the CMS, LHCb, MATHUSLA, FASER, MAPP$^{*}$ and CODEX-b for background free estimates. We take $g_1^{\prime} = 0.001$ and $\frac{m_N}{m_{Z^{\prime}}} = 0.3$ to maximise the phase space for the decays of \zprime. The horizontal band represent the preferred parameter region in which the light neutrinos acquire a mass scale of $m_\nu = V_{lN}^2 m_N$ in the range $10^{-2}~\text{eV} < m_\nu < 0.3$~eV as indicated by oscillation experiments and constrained by direct neutrino mass measurements. It is worth noting the complementarity of different experiments in probing the parameter space of neutrino mass and mixing for a fixed ratio of \zprime to heavy neutrino mass. In order to do this, we first estimate the plot the lab decay length of the heavy neutrinos for two benchmark points in Figure~\ref{fig:histos}. For the two plots, we choose two different neutrino masses of 5 GeV (left) and 15 GeV (right) for a fixed neutrino mixing of $log(V_{\mu N}) = -4.5$. Correspondingly the \zprime\, mass is fixed at 17 and 50 GeV respectively. As the mass difference between heavy neutrino and \zprime\, is larger for latter, in general neutrinos are more boosted i.e. they travel for a longer decay length. We then plot the corresponding decay length distributions after selecting events which pass the geometry of CODEX-b and FASER detectors. We see that for FASER only the most highly boosted events are selected. Finally, we show in vertical lines, the distance at which the detector is located. It is hence clear that both FASER and CODEX-b will have sensitivity for heavy neutrino masses of 5~GeV, however for 15~GeV, neither CODEX-b nor FASER is sensitive. For this mass, if the mixing angle is somewhat smaller, FASER can have sensitivity and this is reflected in Figure~\ref{fig:big}.

Of the parameter space under consideration, at low neutrino masses, FASER 2 probes the largest active-sterile neutrino mixing angle. This sensitivity is further extended by CODEX-b and MAPP$^{*}$ detectors, while the MATHUSLA detector is capable of probing the smallest neutrino mixing angles, reaching in the region interesting for the seesaw mechanisms. At heavier neutrino masses, this reach of MATHUSLA detector is complimented by LHCb and CMS detectors. First, the reach of the experiments is shown for different luminosities. While FASER 2, CMS and MATHUSLA have 3000~fb$^{-1}$, CODEX-b, MAPP$^{*}$ and LHCb have 10 times less luminosity. Second, as discussed previously, the experiments have different volumes and thus are sensitive to different regions of heavy neutrino mass and mixing parameter space. Among the experiments, FASER 2 has one of the smallest geometric reach. Combining this with the luminosity of 150~fb$^{-1}$, the experiment yields no sensitivity over the whole parameter space considered. FASER 2 with 20 times more luminosity, offers more promising prospects. It is also interesting to notice that FASER 2 covers the largest neutrino mixing angle and maximal neutrino mass of 15~GeV. Even if FASER 2 is further away from the IP compared to other detectors, it is not competitive as it is close to beam line and has small radius, effectively small volume. CODEX-b and MAPP$^{*}$ have a reach for smaller neutrino mixing angles for neutrino mass less than 15~GeV. This is essentially because MAPP$^{*}$ has a slightly larger volume compared to CODEX-b. The MATHUSLA detector has the best reach in neutrino mixing angles covering a mixing of up to $10^{-6}$ for all neutrino masses considered. 

While the above optimistic view is possible for a direct projection to higher luminosity, we can also take a pessimistic view by taking the upper limit of background from the non-observation at 2~fb$^{-1}$ \cite{Aaij:2016xmb} and scale it to 300~fb$^{-1}$ at LHCb. Taking an upper limit of 3~background events at 2~fb$^{-1}$ corresponds to a scaled number of 450 background events at 300~fb$^{-1}$ luminosity. Similarly, for CMS we interpret the non-observation of displaced vertex events~\cite{CMS:2014hka, CMS:2015pca} at 20.5~fb$^{-1}$ to yield an upper limit on the mean rate of 3~events \cite{Agashe:2014kda}. We scale this rate up for the 3000~fb$^{-1}$ HL-LHC, which also gives 450 potential background events. Therefore, we set as exclusion if $\chi^{2} = (N_{tot}-N_{B})^2/N_{B}> 3.84$ at 95 $\%$ C.L. \cite{Agashe:2014kda}. In the right panel of Figure~\ref{fig:big}, we show the result of this pessimistic view in which the backgrounds for LHCb and CMS are not assumed to be zero but instead are scaled using existing background limits. The sensitivity at LHCb is clearly reduced, only around $V_{\nu N} \approx 10^{-4}$, while CMS fails to be sensitive for such an assumption. This is mainly due to high $p_T$ cuts for such low mass heavy neutrinos making the number of expected signal events too low to be distinguishable from the expected 450 background events.

\begin{figure}[t!]
	\centering
	\includegraphics[width=\linewidth]{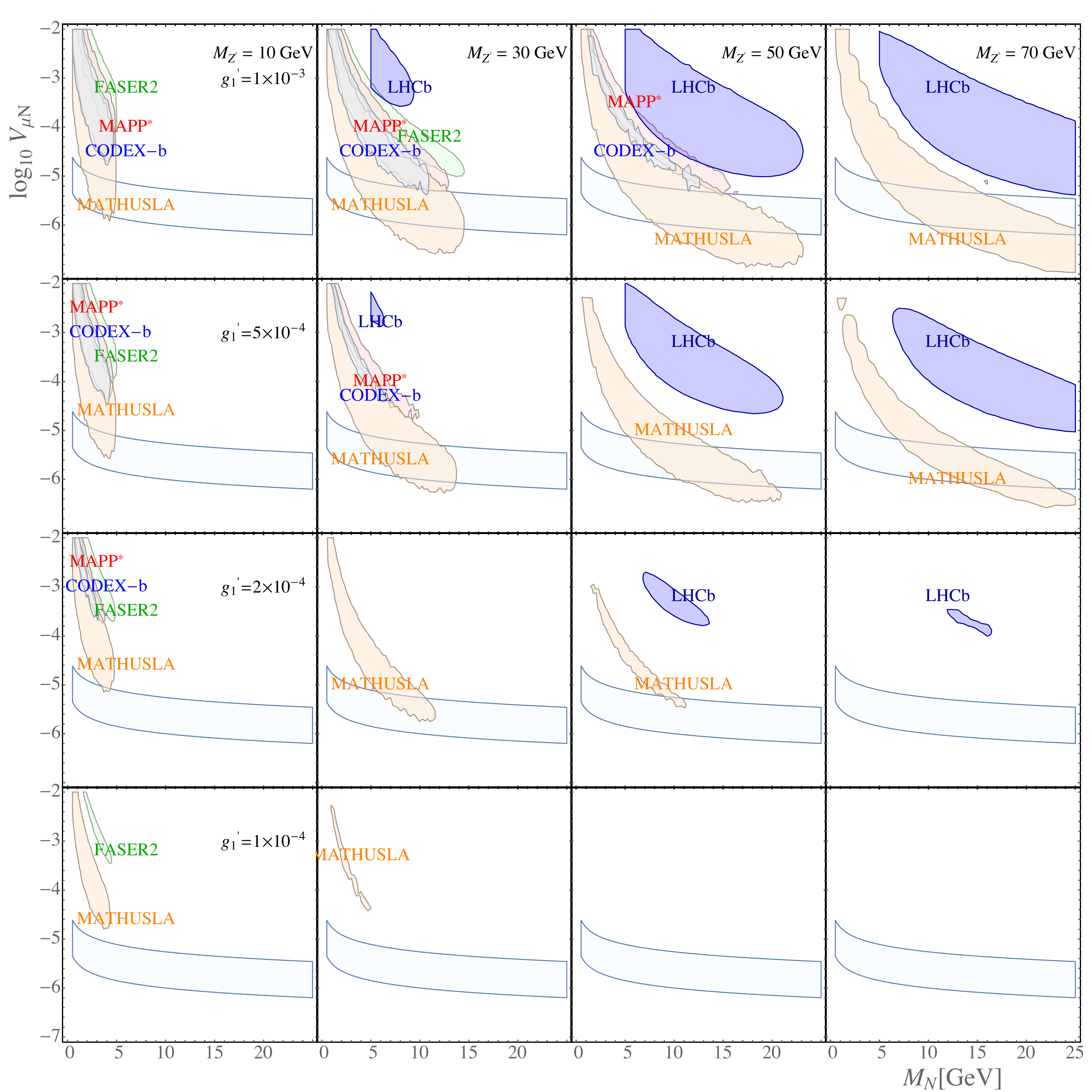}
	\caption{Projected sensitivity of lifetime frontier detectors in the heavy neutrino mass $m_N$ and neutrino mixing $V_{\mu N}$ parameter plane as in Fig.~\ref{fig:big}~(left) but for a series of fixed values of $m_{Z^\prime}$ and $g_1^\prime$ as indicated. The background is assumed to be zero for all detectors.}
	\label{fig:fixmzp}
\end{figure}
In Fig.~\ref{fig:fixmzp} we show the dependence on $m_{Z^\prime}$ and $g_1^\prime$ more explicitly. Instead of fixing $\frac{m_N}{m_{Z^{\prime}}}$, the mass of $Z^{\prime}$ is fixed at 10~GeV in the first column, 30~GeV in the second, 50~GeV in the third and 70~GeV in the fourth. Similarly, $g_1^\prime$ ranges between $10^{-4}$ to $10^{-3}$ from the bottom to the top row. This serves to explore the sensitivities for  different regimes of the $m_{Z^\prime}$ and $m_{N}$ space. The distribution is similar to Fig.~\ref{fig:big} while the sensitivities in $m_{N}$ span differently due to the kinematical threshold for \zprime\, to decay into two heavy neutrinos. Note that due to the $p_{T}(\mu) > 12$~GeV requirement at LHCb, for small $m_{Z^{\prime}}$ such as 10 or 30~GeV, LHCb has no sensitivity, thus LHCb can only probe larger $m_{Z^{\prime}}$ and smaller $V_{\mu N}$ when $m_{Z^{\prime}}$ is about 50~GeV or larger. Additionally, from the selection cuts mentioned in Sec.~\ref{sec:LHCb}, the invariant mass of $\mu j j$ should be bigger than 4.5~GeV, thus we do not have sensitivity for any heavy neutrinos with a mass under 4.5~GeV. This is indicated by the vertical cut-off to the left of the LHCb region. For CMS, as it has even more stringent trigger requirements, we do not get any sensitivity for any neutrino mass. This is consistent with Fig.~\ref{fig:big}~(left) which gives a sensitivity for larger $m_N$ as $m_{Z^\prime}$ is larger than used here. Finally, note that the LHCb sensitivity seems to be better than compared to CMS. This is partly because the signal cross section simulated for LHCb is for the $\mu j j$ final state while for CMS it is for the $\mu\mu\nu$ final state and it also employs a softer $p_{T}$ cut. The gain in the branching ratio due to hadronic decays of the $W$ and a softer $p_{T}$ cut are more significant than the reduction in the luminosity. 

\begin{figure}[t!]
	\centering
	\includegraphics[width=0.49\linewidth]{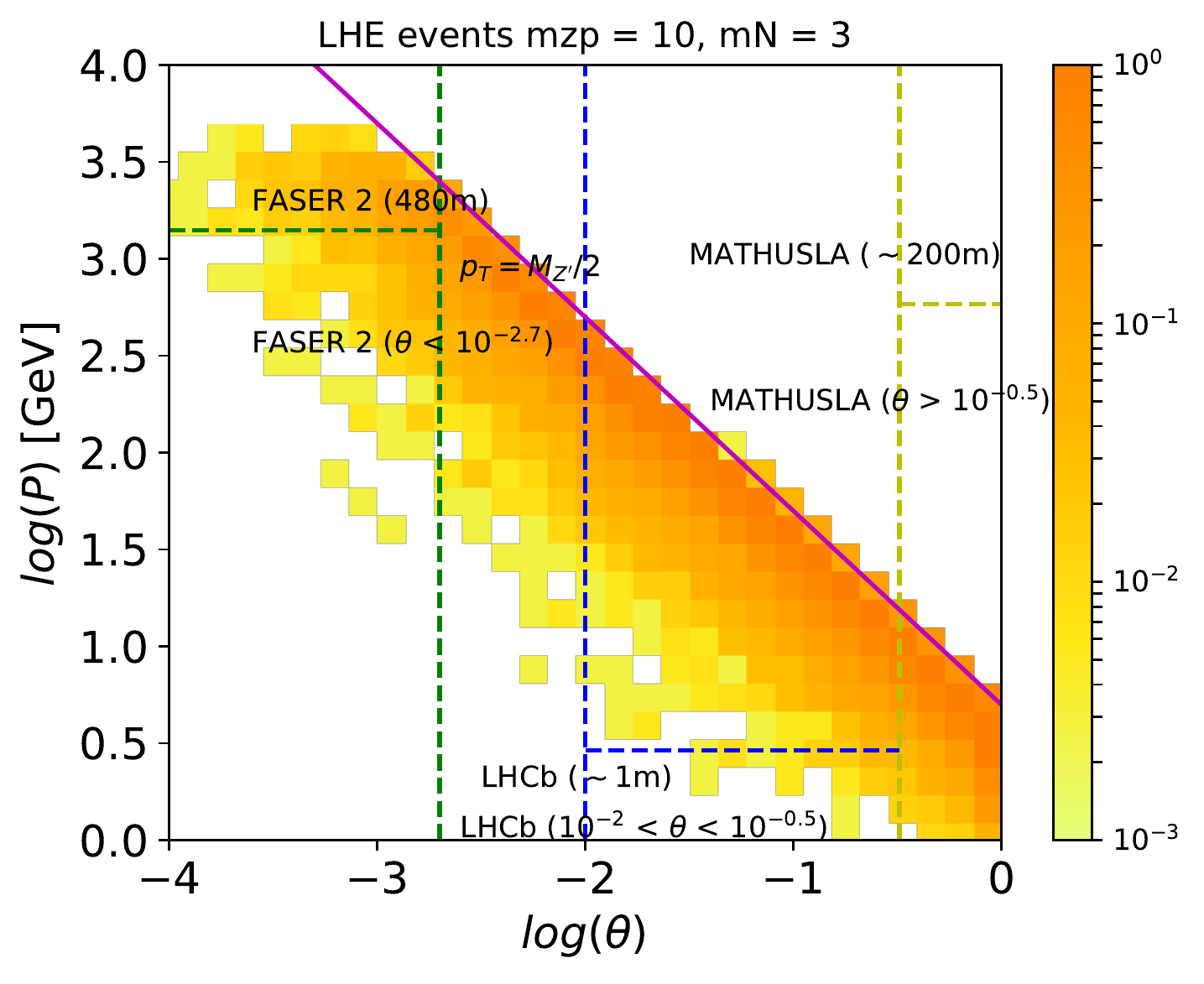}
	\includegraphics[width=0.49\linewidth]{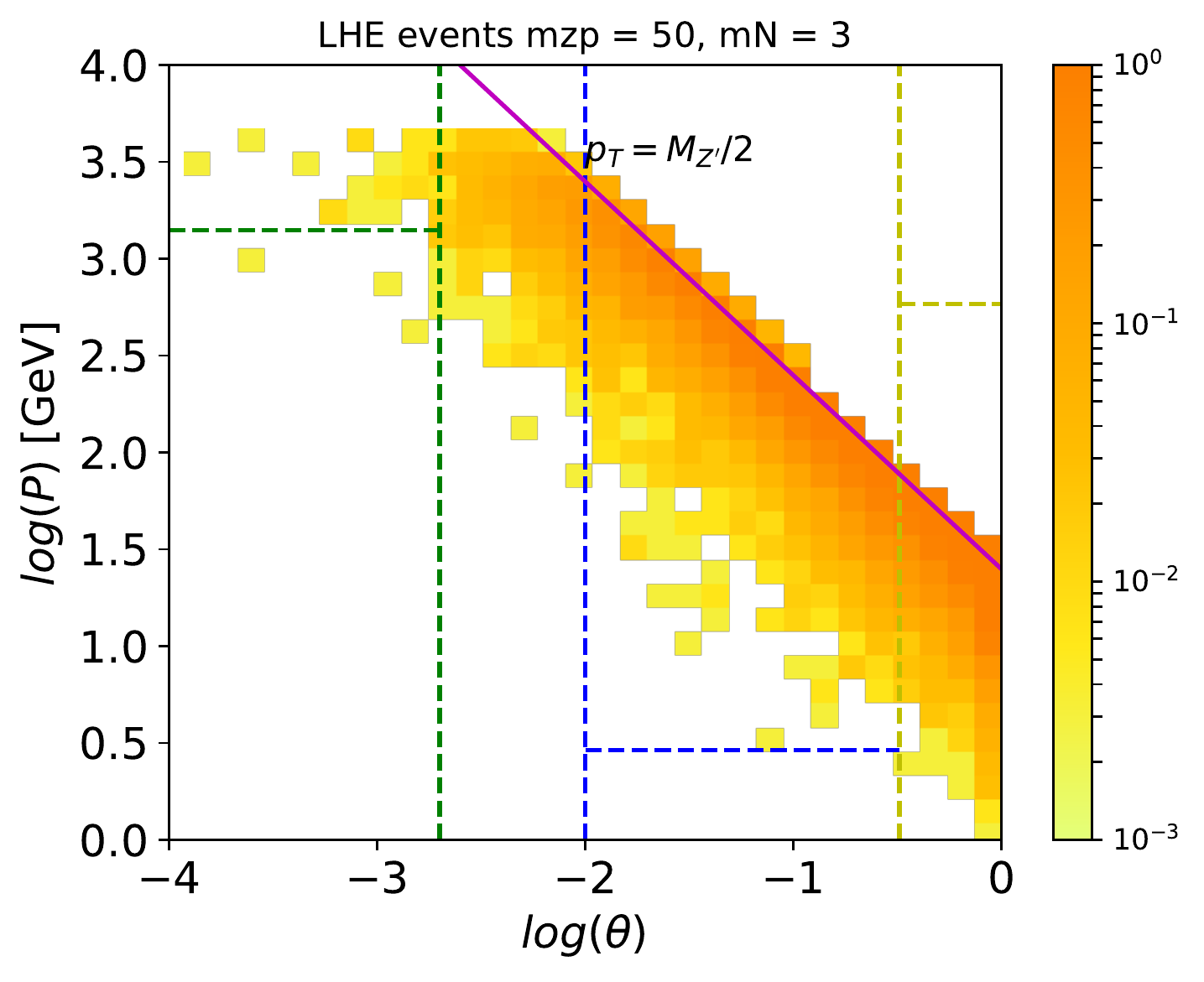}\\
	\includegraphics[width=0.49\linewidth]{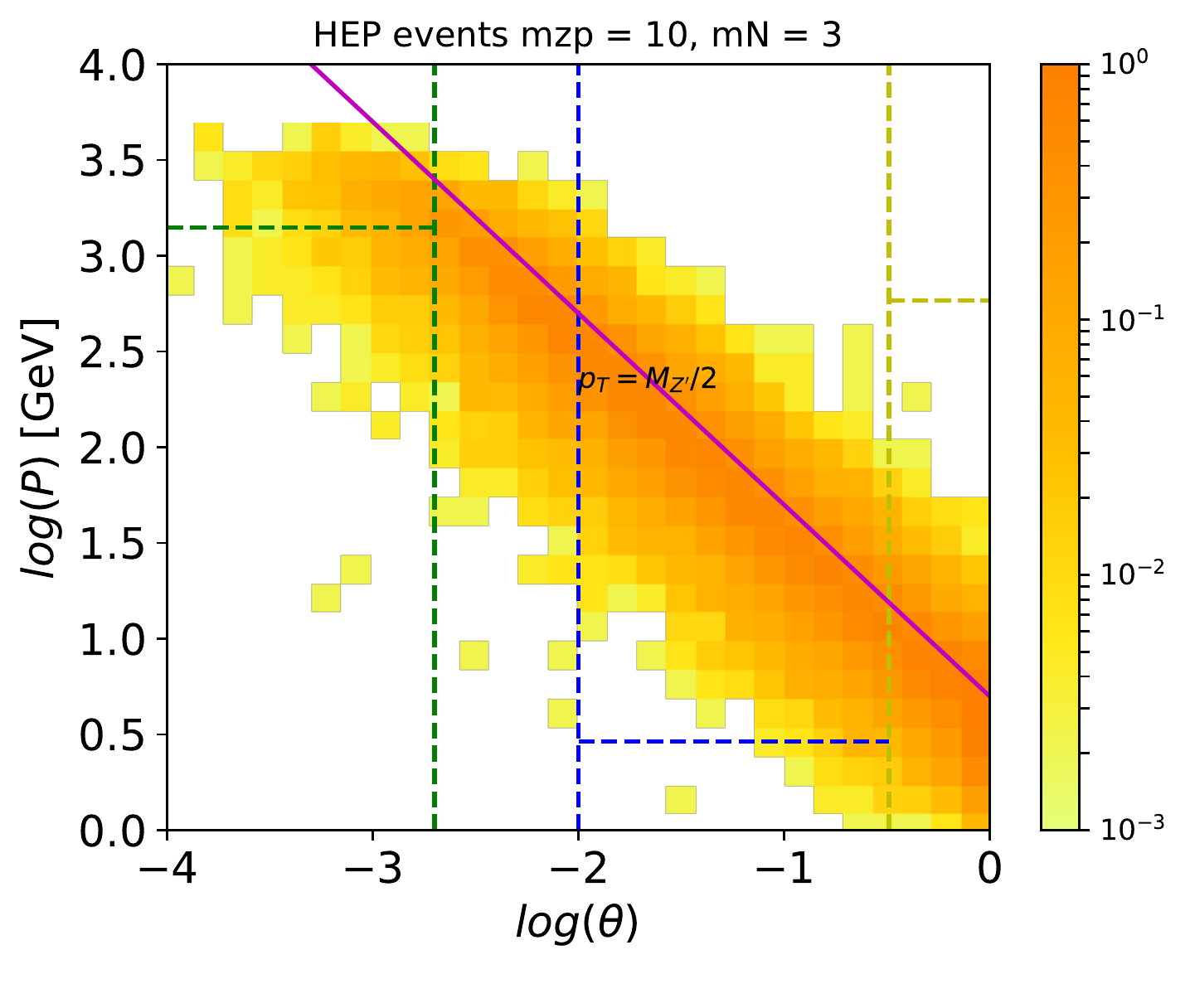}
	\includegraphics[width=0.49\linewidth]{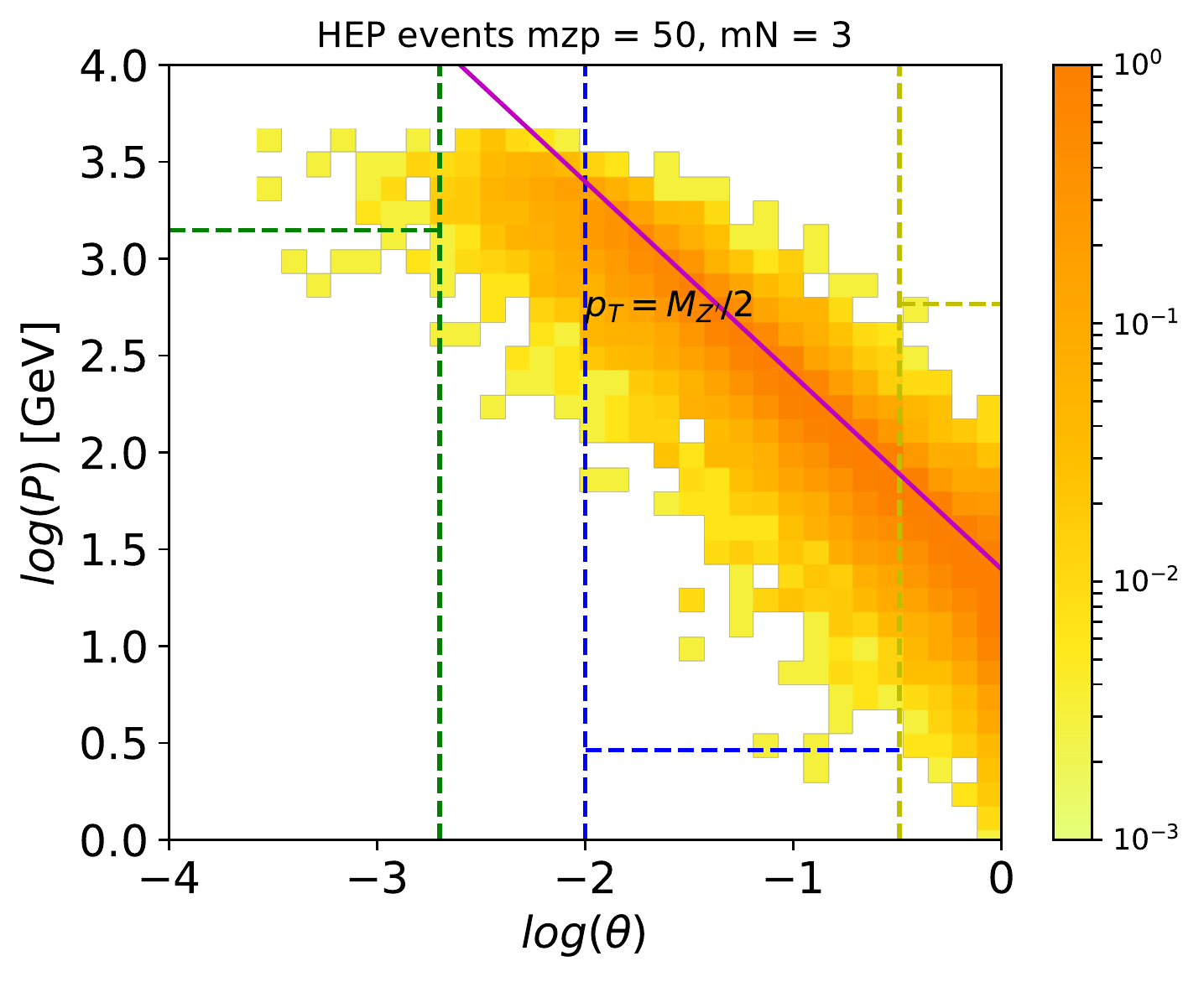}
	\caption{Left: The distribution of the momentum and its angle between the beam line $\theta$ of the heavy neutrino N, with $m_N$ = 3 GeV and $m_{Z^{\prime}}$ = 10 GeV.
		Right: Same but $m_{Z^{\prime}}$ = 50 GeV. Up: The distribution at parton level. Bottom: The distribution at hadronization/shower level. The solid purple line corresponds to $p_T$ = $p \sin \theta$ = $m_{Z^{\prime}}/2$. The horizontal lines indicate the decay length in lab frame when $V_{lN}$ is fixed at $10^{-3}$ for FASER 2 (480m), LHCb ($\sim$1m) and MATHUSLA ($\sim$ 200m) which are taken as representative examples, while the vertical lines show $\theta$ range for FASER 2 ($\theta<10^{-2.7}$), LHCb ($10^{-2.0}<\theta<10^{-0.5}$) and MATHUSLA ($\theta<10^{-0.5}$).}
\label{fig:distro}
\end{figure}
To illustrate the impact of the boost due to different $m_{Z^\prime}$, in Fig.~\ref{fig:distro} we plot the 2D distribution over the total spatial momentum $P$ of the neutrino and the angle $\theta$ it subtends with the beam axis for a fixed small neutrino mass $m_N = 3$~GeV but two different \zprime\ masses, $m_{Z^\prime} = 10$~GeV (left column) and $m_{Z^\prime} = 50$~GeV (right column). We also plot this distribution when simulating the events at parton level (top row) and including the effects of hadronization and showering (bottom row). The horizontal lines indicate momenta corresponding to the decay length in the lab frame for a fixed neutrino mixing of $V_{\mu N} = 10^{-3}$ at 480m, 1m, and 200m. These values correspond approximately to the distance of FASER 2, LHCb, MATHUSLA. This illustrates the different displacements of the heavy neutrino to which the detectors are sensitive. Similarly, the vertical lines show the angle for FASER 2 ($\theta<10^{-2.7}$), LHCb ($10^{-2.0}<\theta<10^{-0.5}$) and MATHUSLA ($\theta<10^{-0.5}$). This demonstrates the different kinematical regions to which the detectors are sensitive. It shows that unlike other detectors, FASER 2 only selects the distribution of heavy neutrinos which are extremely boosted with a very small angle to the beam axis which explains that it only probes the largest mixings in the Fig. \ref{fig:big} and \ref{fig:fixmzp}. Comparing from right to left, larger difference between $m_{Z^{\prime}}$ and $m_{N}$ results in higher average momentum and larger $\theta$. Comparing from bottom to top, the process of hadronization/shower results in a change in the boost distribution of the heavy neutrino due to presence of initial/final state radiation in the process. 

Surveying the above discussions, we conclude that proposed far detectors like MATHUSLA, FASER 2, MAPP$^{*}$ and CODEX-b can reach smaller $m_{N}$ and $V_{\mu N}$. This arises due to a combination of absence of kinematical threshold and large distance from the IP. Among these detectors, although FASER detector is placed very far away from the IP, the momentous boost from the parent particles at the beam direction cancels its advantage, makes it not as competitive as the above detectors for long-lived particles such as heavy neutrinos. CMS and LHCb can only have good sensitivities in the optimistic view for larger $m_{N}$ due to the trigger requirements. 


\section{Conclusion}
\label{conclu}
In this work we have explored the potential to probe the $B-L$ gauge boson $Z^{\prime}$ decaying to two RH neutrinos in the $U(1)_{B-L}$ model. The spontaneous symmetry breaking of the $B-L$ symmetry from the mixing of the SM Higgs field and the additional SM singlet Higgs gives the mass of RH neutrinos. For RH neutrinos with a mass smaller than the half of $m_{Z^{\prime}}$ where we take $m_{N}=0.3 \times m_{Z^{\prime}}$ for simplicity, they can be detected from the $Z^{\prime}$ decays at the beam direction when the mass of $Z^{\prime}$ is tiny. Coincidentally, the searches for dark photons have also be studied at the beam direction especially at LHCb which brings the potential to be recasted as the $B-L$ boson. This recasts give us the bound for the upper bound of the gauge coupling of $B-L$ $g_1^{\prime}$ as a function of the mass of $Z^{\prime}$ as roughly $g_1^{\prime} < 10^{-4}$ for $m_{Z^{\prime}}< 10$ GeV and $g_1^{\prime} < 10^{-3}$ for 10 GeV $<m_{Z^{\prime}}< 100$ GeV. The biggest allowed $g_1^{\prime}$ is than assumed for two cases whether the mass of the $Z^{\prime}$ is smaller or bigger than 10 GeV. After approximate estimation, reinterpreting the data of the dark photon searches at the LHCb shows that the discovery potential for $m_{Z^{\prime}}$ below 10 GeV is almost negligible. The process we have considered in this paper is of an interest when the $B-L$ Higgs is heavy and/or the mixing between $B-L$ Higgs and the SM Higgs is suppressed. 

Considering the final states of the heavy neutrinos with a mass beyond 10 GeV, we can potentially detect them by searching for displaced vertices as the heavy neutrinos can be long-lived when the active-sterile mixings are small. Thus, simulations at LHCb, CMS and proposed detectors MATHUSLA, FASER, MAPP$^{*}$ and CODEX-b are performed. With a focus on the mixing with muons, we show the sensitivities of the active-sterile mixings about $10^{-5}$ for $m_{N}$ smaller than 10 GeV at MAPP$^{*}$ and CODEX-b with 300 fb$^{-1}$ luminosities, while LHCb shows a better sensitivities in the same magnitude for $m_{N} > $ 10 GeV due to the kinematical threshold in need to produce two jets requested by the trigger. CMS can reach $10^{-6}$ for $m_{N} > 25$ GeV to reach high $p_{T}$ cuts, and MATHUSLA can reach a magnitude lower sensitivities due to its long distance from the IP. Although FASER detector is placed very far away from the IP, the momentous boost from the parent particles at the beam direction cancels its advantage, and makes it not as competitive as the above detectors for long-lived particles such as heavy neutrinos. The background of displaced vertices are assumed to be either zero or negligible as the displaced vertices are at least about a meter. However, if we take a pessimistic view, assuming the upper limits of background events, the sensitivities for CMS vanished and becomes much smaller for LHCb. Thus, proposing far detectors can be considered as more sensitive particularly in the pessimistic view. Although, these result are built on the assumption as we choose the biggest $B-L$ couplings allowed which can be excluded by the future searches with higher luminosities.


\acknowledgments

WL acknowledges support via the China Scholarship Council
(Grant CSC No. 2016 08060325). SK is supported by Elise-Richter grant project number V592-N27. We thank Ricardo Cepedello for useful templates on the figures.


\bibliographystyle{JHEP}
\bibliography{F-S-W.bib}
\end{document}